\PassOptionsToPackage{sort&compress}{natbib}
\documentclass[aps,physrev,reprint,onecolumn,11pt,nofootinbib,superscriptaddress,floatfix]{revtex4-2}
\usepackage{amsmath,amssymb}
\usepackage[T1]{fontenc}
\usepackage{newtxtext,newtxmath}
\usepackage{graphicx}
\usepackage{tikz}
\usetikzlibrary{arrows.meta}
\usepackage{float}
\usepackage{microtype}
\usepackage[hidelinks]{hyperref}
\hypersetup{
  pdftitle={A few-parameter flow for lattice phi4 theory fitted by linear least squares},
  pdfauthor={Moxian Qian}
}
\graphicspath{{figures/}}
\newcommand{\dd}{\mathrm d}
\newcommand{\E}{\mathbb E}
\newcommand{\ESS}{\mathrm{ESS}}

\begin{document}

\title{A few-parameter flow for lattice \texorpdfstring{$\phi^4$}{phi4} theory fitted by linear least squares}
\author{Moxian Qian}
\email{mqian@students.uni-mainz.de}
\affiliation{Johannes Gutenberg University Mainz, 55128 Mainz, Germany}

\begin{abstract}
Training a flow-based sampler for a lattice field theory is expensive and is usually repeated for every coupling and volume. We fit the few parameters of a trivializing flow by flow matching. Training then becomes linear least squares on about one thousand Monte Carlo configurations and takes minutes on a CPU. The two peaks of the magnetization distribution in the broken phase are built into the base distribution, and the integrator gives an exact density, so reweighting or an independence Metropolis--Hastings test corrects the samples to the Boltzmann distribution. For the two-dimensional $\phi^4$ theory the corrected results agree with hybrid Monte Carlo within two standard errors at most points. The effective sample size decreases with the volume, most strongly near criticality, where it falls from 0.90 on an $8\times8$ lattice to 0.013 on a $32\times32$ lattice. Reweighting also gives the free energy at a single coupling.
\end{abstract}

\maketitle

\section{Introduction}
\label{sec:intro}

Lattice field theory computes expectation values by sampling configurations from the Boltzmann distribution $e^{-S}/Z$, usually with Markov chain Monte Carlo (MCMC) and in particular with hybrid Monte Carlo (HMC)~\cite{Metropolis:1953am,Hastings:1970aa,Duane:1987de}. Consecutive configurations of a Markov chain are correlated. The number of updates per statistically independent configuration is set by the integrated autocorrelation time. Near a continuous phase transition or the continuum limit it grows as a power of the correlation length (critical slowing down)~\cite{Schaefer:2010hu}. Cluster algorithms such as the Wolff algorithm remove critical slowing down in some spin and scalar models~\cite{Swendsen:1987ce,Wolff:1988uh,Brower:1989mt}.

Generative samplers avoid the chain by producing statistically independent configurations~\cite{Cranmer:2023xbe,Kanwar:2024ujc,Kanwar:2026nlw}. Examples are autoregressive networks~\cite{Wu:2019elz,Qian:2025uvx}, diffusion models~\cite{Wang:2023exq,Zhu:2024kiu,Tan:2026ibs}, stochastic path sampling~\cite{Chen:2026bqw}, and normalizing flows (NFs)~\cite{Rezende:2015ocs,Dinh:2016pgf,Albergo:2019eim,Kanwar:2020xzo,Albergo:2021vyo,Boyda:2020hsi,Abbott:2023thq}. When the sampler provides the exact probability of every configuration, as an NF does, the remaining difference from the Boltzmann distribution is removed exactly by reweighting or by an independence Metropolis--Hastings (IMH) test. The sampler is then judged by the fraction of effectively independent samples it delivers, the effective sample size (ESS). Two limitations are well documented. This fraction decreases with the volume, most strongly near criticality~\cite{DelDebbio:2021qwf,Komijani:2023fzy,Abbott:2022zsh,Singha:2026slc,Iso:2026giq}. Training is expensive. It takes hours to weeks on a GPU~\cite{Gerdes:2022eve,Albergo:2019eim}, up to ten days on 48 A100 GPUs~\cite{Abbott:2024kfc}, and is usually repeated for every coupling and volume~\cite{Albergo:2019eim,Nicoli:2020njz}.

L\"uscher proposed to remove critical slowing down with a trivializing map $F$, which sends Gaussian fields $z$ to configurations $\phi=F(z)$ distributed as $e^{-S}/Z$~\cite{Luscher:2009eq}. The map follows from a flow equation whose velocity is the gradient of a generator, expanded in powers of the flow time with Wilson loops as terms. Only the lowest orders are known. Learned trivializing flows keep this form and train its few coefficients, or a small network in place of the generator~\cite{Bacchio:2022vje,Gerdes:2024rjk,Albandea:2023wgd,Albandea:2023ais}. Analytic bijections follow the same idea for $\phi^4$ theory~\cite{Gerdes:2026rbd}. The training still uses gradient descent on the reverse Kullback--Leibler (KL) divergence. This divergence is estimated from samples of the flow, so every training step integrates the flow and evaluates its density.

We keep this form but fit the coefficients by flow matching, which regresses the velocity of a continuous flow on samples of the target~\cite{Lipman:2023hef,Liu:2022yjp,Albergo:2022iol,Albergo:2025ihk}. For a velocity linear in its coefficients the regression is linear least squares. It needs only configurations of the target, not samples or densities of the flow. Least-squares fits of this kind are used in machine learning for low-dimensional problems and for adapting flows to new distributions~\cite{Ruthotto:2026rf,Khilchuk:2025sb,Ingebrand:2026fm}, and score matching with linear models has the same closed form~\cite{Hyvarinen:2005sm,Hyvarinen:2007sm}.

For the two-dimensional $\phi^4$ theory we combine $K=16$ field functions, motivated by the exact velocity of the free field and by the action, and call the result the few-parameter flow. The two peaks of the magnetization distribution in the broken phase are placed in the base distribution of the spatial mean field (the zero mode). The fit uses about one thousand configurations from HMC with cluster updates. A splitting with strictly monotone substeps turns the fitted flow into a discrete NF with an exact density, and reweighting or IMH corrects its samples to the Boltzmann distribution (Fig.~\ref{fig:scheme}). Cluster updates make reference values and fitting configurations cheap in this model, so it serves as a test bed.

\section{Model and method}
\label{sec:method}

\begin{figure}[!htb]
\centering
\begin{tikzpicture}[font=\footnotesize,
  box/.style={draw,rounded corners=2pt,align=center,inner sep=2.5pt,minimum height=0.85cm,text width=2.25cm},
  arr/.style={-{Stealth[length=4pt]},semithick},
  grp/.style={draw=gray,dashed,rounded corners=4pt}]
\draw[grp] (-1.35,1.15) rectangle (7.0,-1.75);
\node[anchor=north west,text=gray] at (-1.3,1.12) {fit};
\node[box] (data) at (0,0.1) {fitting\\configurations $\phi_1$};
\node[box] (base) at (0,-1.05) {base samples\\$z\sim p_0$};
\node[box] (pair) at (2.85,-0.475) {pair zero modes\\in sorted order};
\node[box] (lsq) at (5.7,-0.475) {linear least\\squares, $16\times16$};
\draw[arr] (data.east) -- (pair.west);
\draw[arr] (base.east) -- (pair.west);
\draw[arr] (pair) -- (lsq);
\draw[grp] (-1.35,-1.95) rectangle (7.0,-5.05);
\node[anchor=north west,text=gray] at (-1.3,-1.98) {sampling};
\node[box] (z) at (0,-3.0) {base samples\\$z\sim p_0$};
\node[box] (flow) at (2.85,-3.0) {splitting with\\monotone substeps};
\node[box] (q) at (5.7,-3.0) {$\phi$, exact $\log q$,\\$w=e^{-S}/q$};
\node[box] (out) at (5.7,-4.35) {reweighting\\or IMH};
\draw[arr] (z) -- (flow);
\draw[arr] (flow) -- (q);
\draw[arr] (q) -- (out);
\draw[arr] (lsq.south) -- node[left]{coefficients $c_n(t_j)$} ++(0,-1.0) -| (flow.north);
\end{tikzpicture}
\caption{The few-parameter flow. In the fit, base samples and fitting configurations are paired in sorted order of their zero modes, and the 16 coefficients are obtained by linear least squares at each interior time node. In the sampling, the flow with these coefficients is integrated by a splitting whose substeps are strictly monotone, which gives configurations with an exact density. Their weights $w$ are used for reweighting or in an IMH chain.}
\label{fig:scheme}
\end{figure}

\subsection{Model and observables}
\label{subsec:model}

The action on a $d$-dimensional periodic lattice with $V=L^d$ sites is
\begin{equation}
S[\phi]=\sum_x\Big[-2\kappa\phi_x\sum_{\mu=1}^{d}\phi_{x+\hat\mu}+(1-2\lambda)\phi_x^2+\lambda\phi_x^4\Big].
\label{eq:action}
\end{equation}
Here $\kappa$ is the hopping parameter and $\lambda$ the quartic coupling. At fixed $\lambda$, increasing $\kappa$ takes the system from a symmetric phase to a broken phase in which the $\mathbb Z_2$ symmetry $\phi\to-\phi$ is spontaneously broken. We take $d=2$ and $\lambda=0.022$, a coupling used in many studies of learned samplers~\cite{Nicoli:2020njz,Nicoli:2023qsl,Caselle:2022acb,Singha:2026slc,Tan:2026ibs}. The magnetization is $M=V^{-1}\sum_x\phi_x$ and the zero-mode coordinate is $y=\sqrt VM$. We measure $\langle|M|\rangle$, the susceptibility $\chi=V\langle M^2\rangle$ (we show $\chi/V=\langle M^2\rangle$), and the Binder cumulant
\begin{equation}
U_4=1-\frac{\langle M^4\rangle}{3\langle M^2\rangle^2},
\label{eq:observables}
\end{equation}
which approaches 0 for a Gaussian magnetization distribution and $2/3$ for two sharp peaks.

\subsection{Exact-density proposal}
\label{subsec:imh}

An NF maps a sample $z$ of a base distribution $p_0$ to a configuration $\phi=F(z)$ by an invertible map $F$, and the change of variables gives the proposal density~\cite{Rezende:2015ocs,Dinh:2016pgf,Albergo:2019eim}
\begin{equation}
\log q(\phi)=\log p_0(z)-\log\bigl|\det\partial F(z)/\partial z\bigr|.
\label{eq:nf}
\end{equation}
With an exact $q$, we draw independent proposals $\phi_i\sim q$ and use them in two ways. The first is an IMH chain. In this chain a proposal replaces the current configuration with probability $\min(1,w'/w)$, where $w=e^{-S[\phi]}/q(\phi)$ is the weight of the current configuration and $w'$ that of the proposal. The chain leaves the target distribution invariant~\cite{Hastings:1970aa,Tierney:1994mc}, and one step of it is one proposal. Like any Markov chain it is characterized by the integrated autocorrelation time $\tau_{\rm int}=\frac12+\sum_{\ell\ge1}\rho(\ell)$ of an observable, where $\rho(\ell)$ is the normalized autocorrelation function at lag $\ell$. $N$ steps are then worth $N_{\rm eff}=N/(2\tau_{\rm int})$ independent samples. We estimate $\tau_{\rm int}$ with the $\Gamma$ method, which truncates the sum at a window chosen from the data~\cite{Wolff:2003sm}. When proposal and target overlap poorly, large weights cause long runs of rejections and $\tau_{\rm int}$ grows~\cite{Bialas:2021bei}.

The second use is reweighting. The proposals are reweighted with the self-normalized estimator $\bar O=\sum_iw_iO_i/\sum_iw_i$. This is importance sampling. We call it reweighting and label it IS in figures and tables. The fraction of effectively independent samples is the ESS
\begin{equation}
\ESS=\frac{\big(\sum_i w_i\big)^2}{N\sum_i w_i^2},
\label{eq:ess}
\end{equation}
where $N$ is the number of proposals. The ESS equals one when $q$ equals the target distribution.

Both uses therefore need a map $F$ whose Jacobian determinant is exactly computable and whose density $q$ is close to $e^{-S}/Z$.

\subsection{Continuous flows and flow matching}
\label{subsec:cnf}

Most lattice NFs build $F$ from a stack of invertible layers, such as affine coupling layers, each chosen so that its Jacobian determinant is cheap~\cite{Dinh:2016pgf,Albergo:2019eim}. A continuous normalizing flow (CNF) is the limit of many infinitesimal layers. It defines $F$ as the solution of $\dd\phi_t/\dd t=v(t,\phi_t)$ from $t=0$ to $1$ with $\phi_0=z$~\cite{Chen:2018wjc}. The velocity needs no invertible structure, but the density
\begin{equation}
\log q(\phi_1)=\log p_0(z)-\int_0^1\nabla\!\cdot v(t,\phi_t)\,\dd t
\label{eq:cnf}
\end{equation}
requires the divergence of $v$ along every trajectory, and lattice CNFs are usually trained by backpropagating through the integral~\cite{Gerdes:2022eve}.

Flow matching replaces this training by a regression~\cite{Lipman:2023hef,Liu:2022yjp}. Pairs $(z,\phi_1)$ are drawn from a joint distribution of $p_0$ and the target, the path is $\phi_t=(1-t)z+t\phi_1$, and the conditional velocity is $u=\phi_1-z$. The loss
\begin{equation}
\mathcal L[v]=\E_{t,(z,\phi_1)}\big\|v(t,\phi_t)-u\big\|^2
\label{eq:cfm}
\end{equation}
is minimized by the marginal velocity $v^\star(t,\phi)=\E[u\mid\phi_t=\phi]$, which transports $p_0$ to the target. Usually $z$ and $\phi_1$ are drawn independently (independent pairing). Pairings based on optimal transport give straighter paths~\cite{Tong:2024dch,Pooladian:2023mf}.

\subsection{Linear velocity and least squares}
\label{subsec:lsq}

If the velocity is restricted to the span of $K$ fixed field functions,
\begin{equation}
v(t,\phi)=\sum_{n=1}^{K}c_n(t)\,f_n(\phi),
\label{eq:intro_velocity}
\end{equation}
the loss~\eqref{eq:cfm} is quadratic in the coefficients. We use 17 time nodes $t_j=j/16$. At the 15 interior nodes we solve
\begin{equation}
c(t_j)=\arg\min_c\ \sum_i\Big\|\sum_n c_nf_n(\phi_{t_j}^{(i)})-u^{(i)}\Big\|^2+\eta\|c\|^2,
\label{eq:lsq}
\end{equation}
where $i$ labels the fitting configurations $\phi_1^{(i)}$ and their paired base samples $z^{(i)}$. Each node is a $K\times K$ linear system. The ridge parameter $\eta$ is chosen on a held-out quarter of the 1024 configurations, the coefficients are fitted on the remaining 768, and they are interpolated linearly between nodes. At the end nodes the coefficients are fixed to the values that are exact for independent pairing (Appendix~\ref{app:endnodes}). The 16 coefficients are fitted together to all $V$ site components of every fitting configuration. The zero-mode functions are constant over the lattice, so their normal equations contain only the spatial mean of the residual, one number per fitting configuration and 768 numbers at each node.

\subsection{Field functions}
\label{subsec:terms}

The $K=16$ functions fall into four groups, each motivated by one part of the problem. The bimodal magnetization of the broken phase motivates three zero-mode terms, and the quartic term of the action motivates three local terms at each site. For a free field the exact flow-matching velocity is linear in the field and diagonal in momentum space, with eigenvalues rational in the lattice momentum as in the preconditioner of Fourier acceleration~\cite{Lipman:2023hef,Albergo:2025ihk,Batrouni:1985jn}. This motivates a Fourier operator on the nonzero modes. Two nearest-neighbor terms account for products of neighboring fields beyond the quadratic coupling. The first is the gradient of $\sum_{\langle xx'\rangle}\phi_x^2\phi_{x'}^2$. The second is the part of the gradient of $\sum_{\langle xx'\rangle}\phi_x^3\phi_{x'}$ that is quadratic in $\phi_x$, with $\phi_x^2$ replaced by the bounded function $r_2$ defined below. With $\psi_x=\phi_x-M$, its Fourier transform $\widehat\psi_k$, and $\widehat k^2=4\sum_\mu\sin^2(k_\mu/2)$, the velocity is
\begin{align}
v_x={}&c_{M1}M+c_{M3}\,r_3(M;s_M)+c_{M5}\,r_5(M;s_M)\nonumber\\
&+\sum_{j=0}^{2}c_{\phi j}\,r_3(\phi_x;2^js_\phi)+[\mathcal G_t(M)\psi]_x\nonumber\\
&+2c_{\rm nn1}\,\phi_x\sum_{x'\sim x}\phi_{x'}^2+3c_{\rm nn2}\,r_2(\phi_x;s_\phi)\sum_{x'\sim x}\phi_{x'},
\label{eq:velocity}
\end{align}
with all coefficients depending on $t$ and $x'\sim x$ denoting nearest neighbors. The saturating functions $r_3(\xi;s)=\xi^3/(1+\xi^2/s^2)$ and $r_5(\xi;s)=\xi^5/(1+\xi^4/s^4)$ are cubic and quintic near the origin and grow linearly for $|\xi|\gg s$. The function $r_2(\xi;s)=\xi^2/(1+\xi^2/s^2)$ is bounded. The scales $s_\phi$ and $s_M$ are the root-mean-square field and magnetization on the interpolating paths, averaged over $t$. The operator $\mathcal G_t(M)$ acts on the nonzero modes with eigenvalues
\begin{align}
\gamma_t(k,M)={}&c_{\gamma0}-c_{\gamma1}\widehat k^2+\big[c_{\gamma2}+c_{\gamma3}H(M)\big]\mathbf 1_{k\in\Lambda_1}\nonumber\\
&+\sum_{j=1}^{4}c_{\omega j}\frac{\omega_j}{\omega_j+\widehat k^2},
\label{eq:rates}
\end{align}
where $\Lambda_1$ is the lowest nonzero momentum shell, $H(M)=M^2/(s_M^2+M^2)$, and $\omega_j\in\{1/16,1/4,1,4\}$. The first line of Eq.~\eqref{eq:velocity} thus contains the three zero-mode terms, the second the three local terms and the Fourier operator with its eight coefficients, and the third the two nearest-neighbor terms ($3+3+8+2=16$). Each coefficient $c_n(t)$ takes one value at each of the 15 interior time nodes, so the flow has 240 fitted coefficients. All 16 functions are odd under $\phi\to-\phi$, so the flow is $\mathbb Z_2$ symmetric. The choice of the local scales is explained in Appendix~\ref{app:basis}.

\subsection{Zero-mode base distribution and pairing}
\label{subsec:zeromode}

In the broken phase the target zero mode $y$ is bimodal, while that of a Gaussian base is unimodal. With independent pairing half of the pairs $(z,\phi_1)$ then have opposite signs of $y$. With a Gaussian base and without pairing, the integration of the fitted flow fails in some runs above $\kappa_c$ (Appendix~\ref{app:extra}). We therefore place the bimodality in the base distribution. Its nonzero modes are standard Gaussian, and its zero mode follows
\begin{equation}
p(y)\propto\exp\!\big[-\alpha(y/y_0)^2-(y/y_0)^4\big],
\label{eq:landau}
\end{equation}
with $\alpha$ and $y_0$ fixed by $\langle y^2\rangle$ and $\langle y^4\rangle$ of the fitting configurations. This is the form of the finite-volume magnetization distribution near the transition~\cite{Binder:1981sa}. Related treatments of the zero mode are a bimodal prior~\cite{Hackett:2021idh} and a learned zero-mode bijection~\cite{Gerdes:2026rbd}. In the pairs of Eq.~\eqref{eq:cfm} the zero modes of base samples and fitting configurations are paired monotonically in $y$ (sorted by value), the one-dimensional case of optimal-transport pairing~\cite{Tong:2024dch,Pooladian:2023mf}. The pairing changes only the fitting path, and sampling draws $z$ directly from $p_0$ (Appendix~\ref{app:landau}).

\subsection{Discretization and exact density}
\label{subsec:integrator}

With the coefficients fixed, the continuous flow still has to be turned into a map whose density is exactly computable. The velocity is integrated with a symmetric (Strang) splitting with step $h=1/32$~\cite{Strang:1968ds}, one substep per group of functions and two for the nearest-neighbor group. The Fourier substep is the explicit scaling $\widehat\psi_k\to e^{h\gamma_t(k,M)}\widehat\psi_k$, whose Jacobian is known in closed form. The zero-mode substep is a one-dimensional equation for $y$, and the local and nearest-neighbor substeps are one-dimensional equations at each site. For the nearest-neighbor terms, one checkerboard color is updated while the other is fixed. These equations are integrated by an implicit midpoint rule with adaptive microsteps, each of which is strictly monotone. The number of microsteps depends only on the coefficients and on the sites held fixed, so the inverse map uses the same subdivision. The discrete map is therefore a bijection for any coefficients, and
\begin{equation}
\log q(F(z))=\log p_0(z)-\sum_\nu\log J_\nu,
\label{eq:density}
\end{equation}
where $F$ is the full discrete map and $J_\nu$ the Jacobian determinant of the $\nu$th substep, is exact up to the tolerance of the midpoint solve (Appendix~\ref{app:integrator}). The sampler is thus again a discrete NF, whose layers are the substeps of the continuous flow, and the divergence integral of Eq.~\eqref{eq:cnf} is never evaluated.

\subsection{Procedure}
\label{subsec:procedure}

The fit and the sampling proceed as follows.

\textit{Fit.}
\begin{enumerate}
\item Generate 1024 fitting configurations by HMC with cluster updates~\cite{Wolff:1988uh,Brower:1989mt} (Appendix~\ref{app:setup}).
\item Fix the zero-mode base distribution, Eq.~\eqref{eq:landau}, from $\langle y^2\rangle$ and $\langle y^4\rangle$ of these configurations (Sec.~\ref{subsec:zeromode}).
\item Draw one base sample per configuration and pair the zero modes in sorted order.
\item Solve the linear least-squares problem for the 16 coefficients at each of 15 time nodes, with the ridge parameter chosen on a held-out quarter (Sec.~\ref{subsec:lsq}).
\end{enumerate}
\textit{Sampling.}
\begin{enumerate}
\item Draw $z$ from the base distribution.
\item Integrate the velocity with 32 splitting steps and accumulate the logarithm of the Jacobian (Sec.~\ref{subsec:integrator}).
\item Compute the weight $w=e^{-S}/q$.
\item Reweight, or run an IMH chain on the proposals.
\end{enumerate}
On four CPU threads, step 1 takes from 40\,s at $L=8$ to 15 minutes at $L=32$, and the least-squares fit 2.5 to 7\,s (Table~\ref{tab:train_time}).

\begin{table}[!htb]
\caption{Cost of the few-parameter flow and of HMC at $\kappa=0.2705$ on four threads of an Intel Xeon 6972P node. The fit times are averages over three seeds. Proposals are generated in batches of 512, and the time is divided by the number of proposals. The HMC time per trajectory is that of 64 chains updated together, divided by 64.}
\label{tab:train_time}
\footnotesize
\begin{ruledtabular}
\begin{tabular}{lrrrrrr}
$L$ & 8 & 12 & 16 & 20 & 24 & 32 \\
\hline
Fitting configurations (s) & 38 & 68 & 138 & 259 & 349 & 903 \\
Least-squares fit (s) & 2.5 & 2.8 & 3.4 & 3.9 & 4.6 & 7.2 \\
Proposal (ms) & 1.5 & 1.5 & 2.0 & 3.6 & 4.5 & 13 \\
HMC trajectory (ms) & 0.015 & 0.023 & 0.033 & 0.047 & 0.058 & 0.090 \\
\end{tabular}
\end{ruledtabular}
\end{table}

\section{Results}
\label{sec:results}

We study $L=8$--32 and $\kappa=0.25$--0.30 around the critical coupling $\kappa_c=0.2705(7)$~\cite{Singha:2026slc}. Reference values come from long HMC chains (Appendix~\ref{app:setup}).

\subsection{Physical observables}
\label{subsec:observables}

\begin{figure}[!htbp]
\centering
\includegraphics[width=\textwidth]{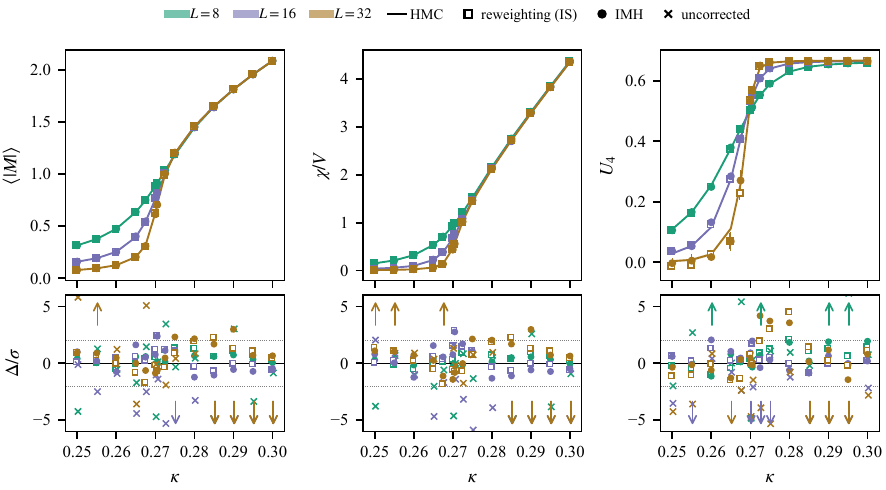}
\caption{$\langle|M|\rangle$, $\chi/V$, and $U_4$ of the two-dimensional $\phi^4$ theory at $\lambda=0.022$ as functions of $\kappa$, with colors distinguishing $L=8$, 16, and 32. Solid lines are the long HMC chains. Squares and circles are reweighting (IS) and IMH with the few-parameter flow, and crosses are its proposals without correction. The narrow panels below each observable show $\Delta/\sigma$, where $\Delta=O-O_{\rm HMC}$ is the deviation from HMC and $\sigma$ the two errors added in quadrature. Dotted lines mark $\pm2$, and arrows mark values beyond $\pm6$. Numerical values for HMC and the few-parameter flow are given in the Supplemental Material~\cite{SM}.}
\label{fig:main}
\end{figure}

Figure~\ref{fig:main} compares the few-parameter flow with HMC. Reweighting and IMH with the few-parameter flow agree with HMC within two standard errors at most points. The tables of the Supplemental Material contain 252 points, each with reweighting and IMH. Of these 504 comparisons, 37 exceed two standard errors. All 37 lie above HMC. Of them 28 are at $L\ge20$ and $0.2725\le\kappa\le0.29$, and four are at $L=16$ and $\kappa_c$. The largest relative differences, up to 4\% in $\langle|M|\rangle$ and 7\% in $\chi/V$, occur just below $\kappa_c$ on the largest lattices. They lie below HMC and within two standard errors. Just above $\kappa_c$ on the largest lattices $U_4$ lies above HMC by up to 4.5 standard errors. This excess does not decrease with more proposals (Appendix~\ref{app:extra}).

\begin{figure}[!htbp]
\centering
\includegraphics[width=0.6\textwidth]{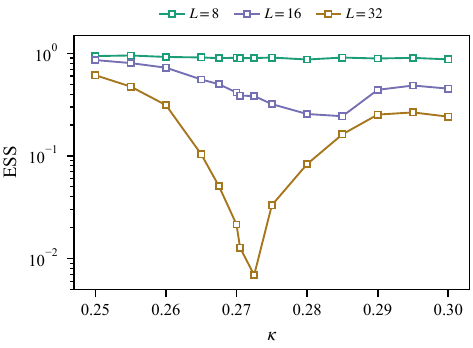}
\caption{ESS of the few-parameter flow as a function of $\kappa$ for $L=8$, 16, and 32.}
\label{fig:ess}
\end{figure}

Without reweighting or IMH, the flow deviates visibly near criticality and in the broken phase. Figure~\ref{fig:ess} shows the ESS. It is smallest near and just above the critical point and decreases with $L$. At $L=16$ it is also low in the broken phase, 0.24 to 0.26 at $\kappa=0.28$ and 0.285. Appendix~\ref{app:conv} compares the ESS with that of convolutional flows trained with reverse KL.

For $\kappa\ge\kappa_c$ the ESS of the few-parameter flow relies on the pairing of the zero modes. Without the pairing it drops at $L=16$ by a factor of 3 to 4 (Appendix~\ref{app:extra}).

\subsection{Volume dependence of the ESS}
\label{subsec:scaling}

Up to $\kappa=0.275$ the ESS decreases with $L$ (Fig.~\ref{fig:ess}). For $\kappa\le0.26$, $-\log\ESS/V$ stays between $0.5\times10^{-3}$ and $1.3\times10^{-3}$ for all $L$, so $-\log\ESS$ grows at most in proportion to the volume. At $\kappa_c$ it grows from $1.7\times10^{-3}$ at $L=8$ to $5.0\times10^{-3}$ at $L=24$ and is $4.2\times10^{-3}$ at $L=32$. For $\kappa\ge0.285$ it lies between 1.3 and $3\times10^{-3}$ at all $L$ except $L=16$ (Supplemental Material).

\subsection{IMH chains and HMC}
\label{subsec:tau}

\begin{figure}[!htbp]
\centering
\includegraphics[width=\textwidth]{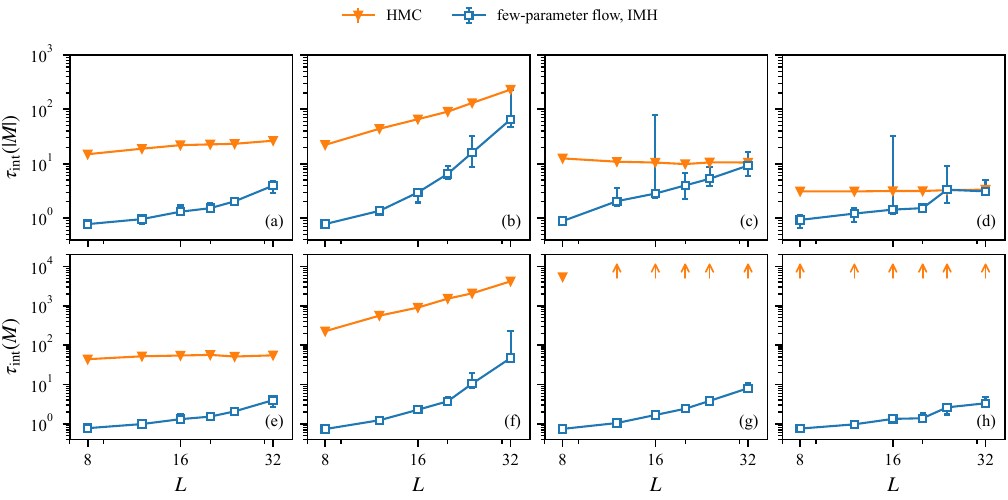}
\caption{Integrated autocorrelation times versus $L$ for $|M|$ [(a)--(d)] and for the signed magnetization $M$ [(e)--(h)] at $\kappa=0.26$ (symmetric phase), $\kappa=0.2705$ (critical point), and $\kappa=0.28$ and 0.30 (broken phase), from left to right. For the few-parameter flow the median and the full range over four independent chains are shown. One step is one proposal for IMH and one trajectory for HMC, whose errors are from the $\Gamma$ method~\cite{Wolff:2003sm}. Arrows mark HMC values of $\tau_{\rm int}(M)$ above the plotted range.}
\label{fig:tau_main}
\end{figure}

The ESS refers to independent proposals. In an IMH chain a low ESS appears as runs of rejections, that is, as autocorrelation. Figure~\ref{fig:tau_main} shows $\tau_{\rm int}$ as a function of $L$. On small lattices the IMH chain of the few-parameter flow decorrelates within about one step at all couplings, while HMC needs up to tens of trajectories for $|M|$. The two are counted in different units, and one proposal costs as much as 60 to 150 HMC trajectories (Table~\ref{tab:train_time}).

With the volume, the IMH autocorrelation grows faster than that of HMC. At $\kappa_c$, $\tau_{\rm int}(|M|)$ of the IMH chain rises from below one at $L=8$ to about 65 at $L=32$, and that of HMC from about 20 to about 230. For comparison, the multiscale sampler of Ref.~\cite{Singha:2026slc} keeps $\tau_{\rm int}(|M|)$ below one up to $L=32$ at the same coupling.

The signed magnetization $M$ shows whether the proposals cover both signs. For the IMH chain $\tau_{\rm int}(M)$ stays close to $\tau_{\rm int}(|M|)$, below 10 in the symmetric and broken phases and about 50 at $\kappa_c$ and $L=32$. For HMC it is set by the tunneling between the two signs and reaches several thousand at $\kappa_c$ and $L=32$.

\subsection{Free energy}
\label{subsec:free_energy}

\begin{figure}[!htbp]
\centering
\includegraphics[width=0.6\textwidth]{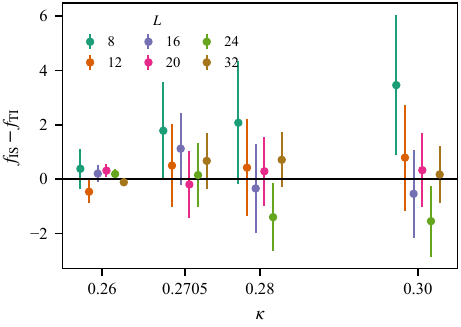}
\caption{Difference $f_{\rm IS}-f_{\rm TI}$ of the free-energy density $f=-\log Z/V$ from reweighting with the few-parameter flow and from thermodynamic integration (TI) with HMC, at $\kappa=0.26$, 0.2705, 0.28, and 0.30 for $L=8$--$32$ (colors). The vertical axis is in units of $10^{-4}$. Error bars combine both statistical errors.}
\label{fig:free_energy}
\end{figure}

Besides expectation values, reweighting gives the partition function directly, $\log Z=\log\frac1N\sum_iw_i$~\cite{Nicoli:2019gun,Nicoli:2020njz}, and hence the free-energy density $f=-\log Z/V$. We compare it with thermodynamic integration (TI) of HMC expectation values. Figure~\ref{fig:free_energy} shows that the two agree. The statistical errors and the computing times of both methods are given in the Supplemental Material.

\section{Discussion}
\label{sec:discussion}

Restricting the flow-matching velocity to a few physically chosen field functions turns training into a linear least-squares fit. For the two-dimensional $\phi^4$ theory the resulting flow agrees with HMC within two standard errors at most points, but its ESS decreases with the volume, most strongly at criticality.

At $\kappa_c$ the ESS of the few-parameter flow is close to that of a diffusion sampler trained without configurations, which has 0.30 at $L=16$ and 0.019 at $L=32$ against our 0.39 and 0.013~\cite{Iso:2026giq}, while our fit uses 1024 HMC configurations. Equivariant continuous flows for the same theory have a velocity linear in a weight tensor for fixed features and are trained by reverse KL~\cite{Gerdes:2022eve}. For fixed features they could be fitted by the same least squares. Their divergence is computed analytically, but their density follows from integrating Eq.~\eqref{eq:cnf} with a fourth-order Runge--Kutta scheme and is exact only up to the integration error. The splitting used here gives the exact density of the discrete map instead.

More fitting configurations do not raise the ESS. In three further fits at $\kappa_c$ with 1024 and with 16\,384 configurations, the mean ESS is 0.408 and 0.416 for $L=16$ and 0.0139 and 0.0135 for $L=32$.

The fit needs equilibrated configurations of the target. In this model cluster updates provide them cheaply. Without such updates it should suffice to equilibrate the configurations within each sign of $M$. The moments $\langle y^2\rangle$ and $\langle y^4\rangle$ do not depend on the sign, and adding the mirrored configurations $-\phi$ restores the balance between the two signs, since the action is symmetric under $\phi\to-\phi$.

For gauge fields we expect the main difficulty to be topological. Vortex numbers and topological charges take integer values and are not smooth functions of the field. Flows whose generator consists of a few Wilson loops exist for gauge theories~\cite{Bacchio:2022vje,Gerdes:2024rjk}, but their parameters are trained with reverse KL. The next step is to fit such generators by least squares and, as for the zero mode here, to build the topological sectors into the base distribution.

\begin{acknowledgments}
The author thanks Shiyang Chen, Cunxi Gong, Biagio Lucini, Zhenkang Lu, Zhuoyi Pang, and Jianhui Zhang for helpful discussions. The computations were performed on the HIMSTER3 cluster of the Helmholtz Institute Mainz and on the CPU and GPU nodes of the IQCB cluster at Johannes Gutenberg University Mainz. Claude Code (Anthropic) helped polish the manuscript and prepare figure and analysis scripts based on the author's code. The author checked all results and is responsible for the content.
\end{acknowledgments}

\section*{Data availability}
The code, the fitted coefficients, and the data behind the figures are available at \url{https://github.com/qxxmax/fewterm-systems}. A notebook there reproduces the main results at $L=8$, 12, and 16.

\appendix

\section{Method details}
\label{app:method}

\subsection{Local and nearest-neighbor terms}
\label{app:basis}

The local terms use three scales $s_\phi$, $2s_\phi$, and $4s_\phi$. With a single scale, $r_3$ grows only linearly for $|\phi_x|\gg s_\phi$. The larger scales extend the cubic growth to larger fields. The two nearest-neighbor terms couple each site to its neighbors. They are integrated in the nearest-neighbor substeps, which update one checkerboard color at a time and keep the Jacobian block-triangular.

\subsection{Zero-mode base distribution}
\label{app:landau}

The kurtosis ratio $\mathcal R(\alpha)=\langle\xi^4\rangle/\langle\xi^2\rangle^2$ of Eq.~\eqref{eq:landau}, with $\xi=y/y_0$, decreases monotonically from 3 ($\alpha\to+\infty$) to 1 ($\alpha\to-\infty$). We solve for $\alpha$ from $\langle y^4\rangle/\langle y^2\rangle^2$ of the fitting configurations and then fix $y_0$ from $\langle y^2\rangle$. When the ratio is not smaller than 2.99, the zero mode is taken to be Gaussian. The variable $y$ is sampled by the inverse cumulative distribution function on a fine grid, and $\log p(y)$ is computed as a piecewise-constant density on the same grid, so $\log p(y)$ is the exact density of the sampled $y$. The grid extends to $\pm9$ standard deviations for the Gaussian case and to where the double-well density has fallen by more than $e^{-600}$. The base probability outside the grid is below $10^{-18}$. The base field is $z=z_\perp+(y/\sqrt V)\mathbf 1$, where $\mathbf 1$ is the field equal to one at every site and $z_\perp$ is a standard Gaussian field with the zero mode removed, so that
\begin{equation}
\log p_0(z)=-\tfrac12\big(\|z\|^2-y^2\big)-\tfrac{V-1}{2}\log2\pi+\log p(y).
\end{equation}
The zero-mode pairing is done once per fit. The $y$ values of the base samples are sorted and assigned to the fitting configurations in the order of their $y$.

\subsection{End nodes}
\label{app:endnodes}
At $t=0$ and 1 the coefficients are fixed such that $v(0,z)=-z$ and $v(1,\phi)=\phi$. For independent pairing and a $\mathbb Z_2$-symmetric target these are the exact values of $v^\star$. With the zero-mode pairing they remain exact for the nonzero modes, because translation invariance gives $\E[\phi_{1,x}-M_1\mid y_1]=0$, where $M_1$ and $y_1$ belong to $\phi_1$. For the zero mode the exact values become $[T(y)-y]/\sqrt V$ at $t=0$ and $[y-T^{-1}(y)]/\sqrt V$ at $t=1$, where $T$ is the monotone map defined by the pairing. We keep $-M$ and $M$ for the zero mode as well. This choice changes the flow only on the first and last time intervals, $[0,1/16]$ and $[15/16,1]$, and the density remains exact.

\subsection{Integrator}
\label{app:integrator}

Each of the 32 steps executes a zero-mode half step, a local half step, two nearest-neighbor half steps, one per checkerboard color, and a full Fourier step, and then the half steps in reverse order. The Fourier step is $\widehat\psi_k\to e^{h\gamma_t(k,M)}\widehat\psi_k$, with log-Jacobian $h\sum_{k\ne0}\gamma_t$. The zero-mode substep is an ordinary differential equation for $y$. The local and nearest-neighbor substeps are one-dimensional equations $\dot\phi_x=g(\phi_x)$ at each site, with the sites of the other color fixed in the nearest-neighbor substeps. They are integrated with the implicit midpoint rule. Each substep is divided into microsteps of size $h_{\rm micro}$ such that $\theta=h_{\rm micro}\,g'/2$ satisfies $|\theta|\le1/4$, using an upper bound on $g'$ that depends only on the coefficients and on the fixed sites. The midpoint equation then has a unique solution, the map is strictly monotone, and each microstep contributes $\log[(1+\theta)/(1-\theta)]$ to $\log J_\nu$, with $g'$ evaluated at the midpoint. The inverse map runs the steps in reverse order with the opposite sign. Each substep integrates one group of terms of Eq.~\eqref{eq:velocity}, and the groups add up to $v$, so the local and nearest-neighbor substeps also change $M$. At finite $h$ the fixed order of the two colors can make $q$ differ between the two sublattices. Reweighting and IMH remove this difference like any other difference between $q$ and the target.

\section{Numerical setup}
\label{app:setup}

Table~\ref{tab:runs} lists the sample sizes.

\begin{table}[!htb]
\caption{Sample sizes.}
\label{tab:runs}
\footnotesize
\begin{ruledtabular}
\begin{tabular}{p{0.4\columnwidth}p{0.52\columnwidth}}
HMC reference chain per point & $2\times10^6$ trajectories after $10^4$ of warm-up\\
Fitting configurations per seed & 1024 chains of 4096 updates\\
Seeds & 3 at $\kappa\in\{0.26,0.2705,0.28,0.30\}$, 1 elsewhere\\
Proposals per seed & 16\,384\\
IMH chains for $\tau_{\rm int}$ & 4 at each $\kappa\in\{0.26,0.2705,0.28,0.30\}$, one per seed ($1.6\times10^4$ steps) and one from a further fit ($5\times10^4$ steps)\\
Convolutional flow, training & 2500 steps of batch size 64, 3 seeds\\
Convolutional flow, sampling & $6.6\times10^4$ proposals and 4 IMH chains of $6.6\times10^4$ steps per seed\\
Large proposal sets (Appendix~\ref{app:extra}) & $1.3\times10^5$ proposals per point\\
Thermodynamic integration & 121 nodes with $2\times10^5$ trajectories each\\
\end{tabular}
\end{ruledtabular}
\end{table}

\paragraph{Reference chains.} Pure HMC uses trajectory length 1 with 10 leapfrog steps of size $\epsilon=0.1$. These chains give the reference values of the observables and the HMC autocorrelation times.

\paragraph{Fitting configurations.} In HMC with cluster updates one HMC trajectory counts as one update. Every fifth update is followed by one Wolff single-cluster update~\cite{Wolff:1988uh,Brower:1989mt} and one checkerboard Metropolis sweep. The fitting configurations are the final states of independent chains. The leapfrog step is $\epsilon=1/\lceil1/\epsilon_0\rceil$ with $\epsilon_0=0.1\min(1,\sqrt{16/L})$, which keeps the acceptance above 0.95.

\paragraph{Fit and integration.} After normalizing each column of the design matrix, we choose the ridge parameter at each node from $\eta\in\{10^{-8},10^{-6},10^{-4},10^{-3},10^{-2},10^{-1}\}$ by the mean-square error on the held-out quarter of the 1024 configurations. The coefficients are those fitted on the other 768. All computations use double precision. The implicit-midpoint equations are solved by Newton iteration to a residual below $10^{-13}$. The double-well zero-mode distribution is tabulated on a grid of $4\times10^5$ points.

\paragraph{Seeds and chains.} Each seed generates its own fitting configurations, fit, and proposals. The IMH estimates of Fig.~\ref{fig:main} come from the chain obtained by accepting or rejecting the proposals in sequence. For the autocorrelation times, each independent flow gives one IMH chain, and Fig.~\ref{fig:tau_main} shows the median and the full range over these chains.

\paragraph{Thermodynamic integration.} The nodes have spacing 0.0025 on $\kappa\in[0,0.30]$. At each node an HMC chain with the reference settings gives $\partial_\kappa\log Z=2\langle\sum_{x,\mu}\phi_x\phi_{x+\hat\mu}\rangle$ with a $\Gamma$-method error. The derivative is integrated after cubic-spline interpolation, and $\log Z(0)$ follows from a single-site integral. The flow estimate of $\log Z$ is averaged over three seeds.

\paragraph{Errors.} The error bars of $-\log\ESS/V$ in Fig.~\ref{fig:akappa} are the standard errors over seeds. All errors are one standard deviation. Errors of chain averages and of $\tau_{\rm int}$ come from the $\Gamma$ method (pyerrors, window parameter $S_\tau=2$)~\cite{Wolff:2003sm,Joswig:2022qfe}, and errors of reweighted estimates from a bootstrap with 200 resamples. Independent seeds are combined by the inverse-variance weighted mean. If $\chi^2/\mathrm{dof}>1$, where dof is the number of seeds minus one, the error is multiplied by $\sqrt{\chi^2/\mathrm{dof}}$.

\section{Additional results}
\label{app:extra}

\paragraph{HMC autocorrelation times.} Table~\ref{tab:tau} lists the HMC autocorrelation times at $\kappa_c$. At $L=8$, 16, and 32 they agree with the values 22.9(3), 64.6(13), and 228(9) of Ref.~\cite{Singha:2026slc}.

\begin{table}[!htb]
\caption{Integrated autocorrelation times of HMC at $\kappa_c=0.2705$ for $|M|$ and for the signed magnetization $M$.}
\label{tab:tau}
\begin{ruledtabular}
\begin{tabular}{lcccccc}
$L$ & 8 & 12 & 16 & 20 & 24 & 32 \\
\hline
$|M|$ & 22.2(5) & 43.9(14) & 65.8(24) & 90.5(39) & 131(7) & 230(15) \\
$M$ & 226(14) & 560(53) & 900(100) & 1510(210) & 2070(330) & 4180(880) \\
\end{tabular}
\end{ruledtabular}
\end{table}

\paragraph{Binder cumulant just above $\kappa_c$.} At $\kappa=0.28$ and $L\ge20$, $U_4$ exceeds HMC by 2 to $4\times10^{-4}$, up to 4.5 standard errors. This excess does not decrease when the number of proposals is raised from $1.6\times10^4$ to $1.3\times10^5$. At $L=20$ and 32 the largest single weight carries less than 1\% of the total weight in each of four independent fits. The excess is accompanied by a deficit on the inner side of the two peaks of the magnetization distribution. With $M_0$ the median of $|M|$ in HMC, at $L=20$ the probability of $|M|<0.8M_0$ is $6.2\times10^{-3}$ in HMC and 2.7 to $5.3\times10^{-3}$ after reweighting. Removing these configurations from the HMC chain raises its $U_4$ by $3.5\times10^{-4}$. At $\kappa=0.2725$ and 0.275 and $L\ge20$ the deviations of $U_4$ reach 1 to $7\times10^{-3}$, up to 4 standard errors. These points have a single seed and an ESS of at most 0.2.

\begin{figure}[!htb]
\centering
\includegraphics[width=0.6\textwidth]{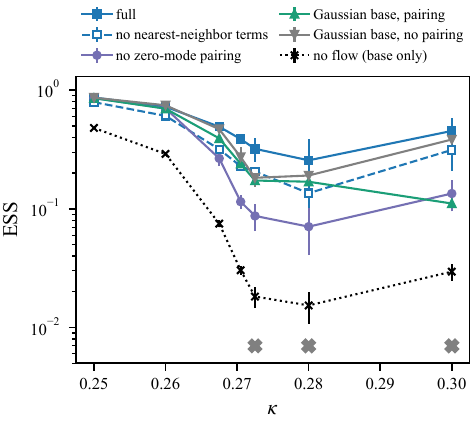}
\caption{ESS at $L=16$ of the full flow, of variants with components removed, and of the free-field proposal with the power spectrum of the fitting configurations and the zero-mode base distribution (no flow). Crosses at the bottom mark couplings at which the integration with a Gaussian base and without pairing failed for at least one of three runs. The points show the remaining runs. All variants, including the full flow, come from fits separate from those of Fig.~\ref{fig:ess}.}
\label{fig:ablation}
\end{figure}

\paragraph{Role of the components.} Figure~\ref{fig:ablation} compares the full flow at $L=16$ with variants in which components are removed. A free-field proposal with the measured power spectrum and the zero-mode base distribution, without the flow, has an ESS about an order of magnitude below the flow at and above $\kappa_c$, so the ESS comes from the flow. In the symmetric phase all flow variants perform alike. From $\kappa_c$ on, the full flow has the highest ESS. The double-well base helps only together with the pairing. Without the pairing its ESS drops by a factor of 3 to 4, below that of a Gaussian base. With a Gaussian base and pairing the ESS is 1.5 to 4 times below that of the full flow. With a Gaussian base and without pairing the integration fails in one of three runs at $\kappa=0.2725$ and in two of three at 0.28 and 0.30, and the remaining runs lie 1.2 to 1.8 times below the full flow. Removing the two nearest-neighbor terms lowers the ESS by a factor of 1.2 to 1.9.

\section{Convolutional flows}
\label{app:conv}

As a reference we train convolutional NFs by reverse KL, which uses no configurations of the target. In the standard setting the flow has 16 affine coupling layers with checkerboard masks and convolutional networks of two hidden layers with 8 channels and kernel size 3, the configuration of the tutorial of Ref.~\cite{Albergo:2021vyo}. To respect the $\mathbb Z_2$ symmetry, its density is the mixture $q_{\rm mix}(\phi)=\frac12[q(\phi)+q(-\phi)]$. It is trained on a CPU with the Adam optimizer~\cite{Kingma:2014vow} at learning rate $10^{-3}$, which takes 390\,s at $L=8$ and 1060\,s at $L=16$ on two threads. We also train it eight times longer, for $2\times10^4$ steps, and a wider version with 32 channels in both hidden layers and $1.6\times10^5$ instead of $1.3\times10^4$ parameters for the same number of steps.

\begin{figure}[!htb]
\centering
\includegraphics[width=0.6\textwidth]{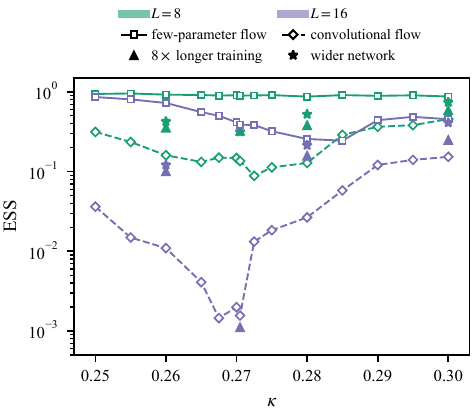}
\caption{ESS of the few-parameter flow (squares, solid lines) and of convolutional flows trained with reverse KL for $L=8$ and 16, in the standard setting (diamonds, dashed lines), trained eight times longer (triangles), and with a wider network (stars).}
\label{fig:conv}
\end{figure}

Figure~\ref{fig:conv} and Table~\ref{tab:convlong} compare the ESS. In the standard setting it is lower than that of the few-parameter flow at every point, by factors of 2 to 10 at $L=8$ and 3 to 350 at $L=16$, with the largest factors near $\kappa_c$. At $L=16$ longer training raises the ESS away from $\kappa_c$ but not at $\kappa_c$, where it stays near $10^{-3}$. The wider network comes within 20\% of the few-parameter flow in the broken phase at $L=16$. In the standard setting at $L=16$, $\tau_{\rm int}(|M|)$ of the IMH chain is about 100 at $\kappa=0.26$, a lower bound (Supplemental Material), and about 200 at $\kappa_c$. These are medians over 12 chains, and the single chain shown in the Supplemental Material gives about 85 at $\kappa_c$.

\begin{table}[!htb]
\caption{ESS of the few-parameter flow and of the convolutional flow in the standard setting, trained eight times longer, and with a wider network. The standard setting and the few-parameter flow are averages over three seeds, the other two columns single seeds. The dash marks a training run that did not finish within 8 hours on two CPU threads.}
\label{tab:convlong}
\begin{ruledtabular}
\begin{tabular}{llllll}
$L$ & $\kappa$ & few-parameter & standard & $8\times$ longer & wider \\
\hline
8 & 0.26 & 0.922 & 0.160 & 0.354 & 0.425 \\
8 & 0.2705 & 0.897 & 0.136 & 0.322 & 0.315 \\
8 & 0.28 & 0.871 & 0.128 & 0.379 & 0.523 \\
8 & 0.30 & 0.873 & 0.455 & 0.586 & 0.740 \\
16 & 0.26 & 0.724 & 0.0109 & 0.100 & 0.121 \\
16 & 0.2705 & 0.387 & 0.0016 & 0.0011 & -- \\
16 & 0.28 & 0.256 & 0.027 & 0.157 & 0.214 \\
16 & 0.30 & 0.453 & 0.153 & 0.247 & 0.410 \\
\end{tabular}
\end{ruledtabular}
\end{table}

\bibliography{references}

\clearpage
\onecolumngrid
\setcounter{section}{0}\setcounter{figure}{0}\setcounter{table}{0}\setcounter{equation}{0}
\renewcommand{\thesection}{S\arabic{section}}
\renewcommand{\thefigure}{S\arabic{figure}}
\renewcommand{\thetable}{S\arabic{table}}
\renewcommand{\theequation}{S\arabic{equation}}
\renewcommand{\theHfigure}{S\arabic{figure}}\renewcommand{\theHtable}{S\arabic{table}}\renewcommand{\theHequation}{S\arabic{equation}}
\renewcommand{\appendixname}{}
\begin{center}\textbf{\large Supplemental Material}\end{center}
\label{sm:start}

The Supplemental Material contains $-\log\ESS/V$ for all volumes, the convergence of the autocorrelation times, the cost and errors of the free energy, and the numerical values of the points in Fig.~\ref{fig:main} for all six volumes.

\section*{S1\texorpdfstring{\quad}{ } Volume dependence of the ESS}
\label{sm:scaling}

\begin{figure}[H]
\centering
\includegraphics[width=0.5\textwidth]{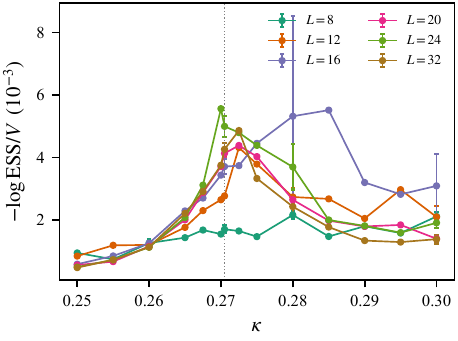}
\caption{$-\log\ESS/V$ of the few-parameter flow versus $\kappa$ for $L=8$--32. Error bars are standard errors over seeds where several seeds exist. The dotted line marks $\kappa_c$.}
\label{fig:akappa}
\end{figure}

Figure~\ref{fig:akappa} shows $-\log\ESS/V$ for all six volumes. The values quoted in Sec.~\ref{subsec:scaling} are read from it.

\section*{S2\texorpdfstring{\quad}{ } Convergence of the autocorrelation times}
\label{sm:acf}

\begin{figure}[H]
\centering
\includegraphics[width=0.85\textwidth]{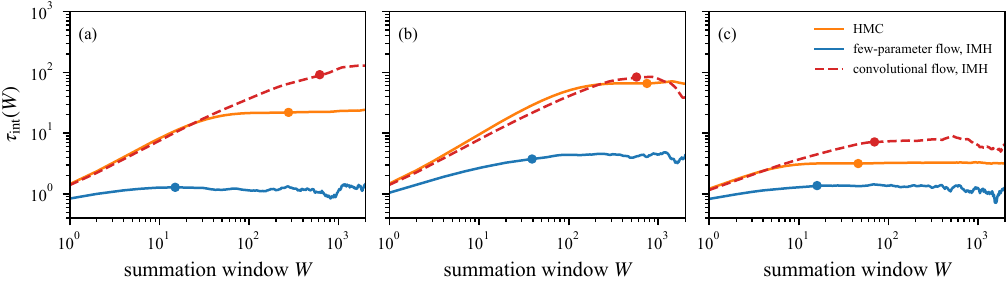}
\caption{Integrated autocorrelation time $\tau_{\rm int}(W)=\frac12+\sum_{\ell=1}^{W}\rho(\ell)$ of $|M|$ at $L=16$ as a function of the summation window $W$, in (a) the symmetric phase ($\kappa=0.26$), (b) at the critical point ($\kappa=0.2705$), and (c) in the broken phase ($\kappa=0.30$), for single chains of HMC ($2\times10^6$ trajectories) and of the IMH chains of the few-parameter flow ($5\times10^4$ steps) and of the convolutional flow ($6.6\times10^4$ steps). Dots mark the window chosen by the $\Gamma$ method and the resulting $\tau_{\rm int}$.}
\label{fig:acf}
\end{figure}

Figure~\ref{fig:acf} shows how $\tau_{\rm int}$ of $|M|$ at $L=16$ builds up when $\rho(\ell)$ is summed up to a window $W$. For HMC the sum reaches a plateau before the window chosen by the $\Gamma$ method. For the IMH chain of the few-parameter flow it does so in the symmetric and broken phases, while at $\kappa_c$ the plateau lies about 17\% above the chosen value. For the IMH chain of the convolutional flow in the symmetric phase the sum is still growing at that window, and its $\tau_{\rm int}$ is a lower bound.

\section*{S3\texorpdfstring{\quad}{ } Cost and errors of the free energy}
\label{sm:timing}

The flow estimate of $\log Z$ from three seeds, including the fit, takes 5 to 60 minutes for $L=8$ to 32 on four CPU threads. TI over $\kappa\in[0,0.30]$ takes 3 to 4.5 hours with single HMC chains, or about 6 to 36 minutes with 64 chains updated together, and gives $f(\kappa)$ on the whole interval. The statistical error of $f$ is $6\times10^{-6}$ to $7\times10^{-5}$ for the flow and $10^{-5}$ to $3\times10^{-4}$ for TI.

\section*{S4\texorpdfstring{\quad}{ } Numerical values}
\label{sm:tables}

Tables~\ref{tab:L8}--\ref{tab:L32} give the numerical values of the points in Fig.~\ref{fig:main} and the same quantities for $L=12$, 20, and 24. HMC is one chain of $2\times10^6$ trajectories (errors from the $\Gamma$ method). IS and IMH are the reweighting and IMH estimates of the few-parameter flow. Numbers in parentheses are one standard deviation on the last digits. At $\kappa\in\{0.26,0.2705,0.28,0.30\}$ three seeds are combined, the observables by the inverse-variance weighted mean and the ESS by the plain mean. The other points have one seed.

\begin{table}[H]
\caption{Observables at $L=8$.}
\label{tab:L8}
\footnotesize
\begin{ruledtabular}
\begin{tabular}{lrrrrrrrrrr}
& & \multicolumn{3}{c}{$\langle|M|\rangle$} & \multicolumn{3}{c}{$\chi/V$} & \multicolumn{3}{c}{$U_4$}\\
$\kappa$ & ESS & HMC & IS & IMH & HMC & IS & IMH & HMC & IS & IMH\\
\hline
0.25 & 0.941 & 0.31523(65) & 0.3156(18) & 0.3161(21) & 0.15120(57) & 0.1513(16) & 0.1518(19) & 0.1090(21) & 0.1072(92) & 0.106(13)\\
0.255 & 0.954 & 0.37658(87) & 0.3766(20) & 0.3767(24) & 0.21117(86) & 0.2110(21) & 0.2113(24) & 0.1657(20) & 0.1622(76) & 0.1660(83)\\
0.26 & 0.922 & 0.4741(12) & 0.4731(15) & 0.4740(19) & 0.3211(14) & 0.3199(19) & 0.3206(23) & 0.2546(19) & 0.2503(38) & 0.2491(44)\\
0.265 & 0.912 & 0.6351(16) & 0.6350(29) & 0.6373(34) & 0.5350(21) & 0.5348(42) & 0.5372(48) & 0.3759(16) & 0.3742(37) & 0.3795(41)\\
0.2675 & 0.898 & 0.7482(18) & 0.7496(32) & 0.7507(39) & 0.7065(26) & 0.7075(51) & 0.7100(61) & 0.4407(13) & 0.4412(31) & 0.4412(36)\\
0.27 & 0.906 & 0.8848(18) & 0.8867(29) & 0.8843(36) & 0.9342(29) & 0.9362(51) & 0.9323(61) & 0.5021(10) & 0.5030(20) & 0.5017(23)\\
0.2705 & 0.897 & 0.9142(18) & 0.9139(21) & 0.9146(22) & 0.9868(30) & 0.9849(32) & 0.9858(39) & 0.51309(99) & 0.5129(12) & 0.5125(13)\\
0.2725 & 0.900 & 1.0335(18) & 1.0380(32) & 1.0340(38) & 1.2101(31) & 1.2167(62) & 1.2092(70) & 0.55185(78) & 0.5534(15) & 0.5533(18)\\
0.275 & 0.910 & 1.1842(16) & 1.1885(26) & 1.1885(34) & 1.5220(30) & 1.5289(56) & 1.5300(72) & 0.58920(54) & 0.59064(98) & 0.5899(13)\\
0.28 & 0.871 & 1.44586(95) & 1.4480(14) & 1.4473(19) & 2.1621(23) & 2.1653(36) & 2.1628(43) & 0.63023(21) & 0.63102(31) & 0.63103(38)\\
0.285 & 0.910 & 1.64646(57) & 1.6477(18) & 1.6478(22) & 2.7574(17) & 2.7601(58) & 2.7610(71) & 0.646144(85) & 0.64642(24) & 0.64621(31)\\
0.29 & 0.891 & 1.81116(41) & 1.8130(16) & 1.8126(19) & 3.3153(14) & 3.3212(56) & 3.3192(68) & 0.653243(38) & 0.65350(19) & 0.65366(22)\\
0.295 & 0.904 & 1.95599(33) & 1.9571(14) & 1.9566(17) & 3.8545(13) & 3.8584(56) & 3.8566(65) & 0.657085(23) & 0.65717(13) & 0.65706(15)\\
0.30 & 0.873 & 2.08713(27) & 2.08747(92) & 2.0872(15) & 4.3804(11) & 4.3814(37) & 4.3801(62) & 0.659438(15) & 0.659531(62) & 0.659570(67)\\
\end{tabular}
\end{ruledtabular}
\end{table}

\begin{table}[H]
\caption{Observables at $L=12$.}
\label{tab:L12}
\footnotesize
\begin{ruledtabular}
\begin{tabular}{lrrrrrrrrrr}
& & \multicolumn{3}{c}{$\langle|M|\rangle$} & \multicolumn{3}{c}{$\chi/V$} & \multicolumn{3}{c}{$U_4$}\\
$\kappa$ & ESS & HMC & IS & IMH & HMC & IS & IMH & HMC & IS & IMH\\
\hline
0.25 & 0.885 & 0.21164(46) & 0.2140(15) & 0.2145(17) & 0.06926(28) & 0.07091(97) & 0.0711(11) & 0.0568(26) & 0.049(13) & 0.052(15)\\
0.255 & 0.842 & 0.25604(65) & 0.2585(19) & 0.2578(22) & 0.10007(47) & 0.1023(15) & 0.1017(17) & 0.0972(27) & 0.090(13) & 0.100(13)\\
0.26 & 0.841 & 0.3348(10) & 0.3350(15) & 0.3345(15) & 0.16575(87) & 0.1662(15) & 0.1664(14) & 0.1805(26) & 0.1796(44) & 0.1708(63)\\
0.265 & 0.776 & 0.4962(17) & 0.4979(29) & 0.4986(37) & 0.3354(18) & 0.3384(35) & 0.3404(46) & 0.3372(22) & 0.3318(47) & 0.3295(59)\\
0.2675 & 0.718 & 0.6336(20) & 0.6346(32) & 0.6328(40) & 0.5100(25) & 0.5116(43) & 0.5103(54) & 0.4347(18) & 0.4336(31) & 0.4309(43)\\
0.27 & 0.683 & 0.8155(22) & 0.8174(33) & 0.8210(42) & 0.7778(31) & 0.7808(52) & 0.7868(64) & 0.5242(13) & 0.5244(20) & 0.5268(26)\\
0.2705 & 0.671 & 0.8552(22) & 0.8549(19) & 0.8554(25) & 0.8414(32) & 0.8423(29) & 0.8430(40) & 0.5394(12) & 0.5387(11) & 0.5382(14)\\
0.2725 & 0.537 & 1.0158(18) & 1.0194(41) & 1.0195(56) & 1.1230(30) & 1.1284(70) & 1.1287(95) & 0.58850(74) & 0.5893(19) & 0.5884(24)\\
0.275 & 0.580 & 1.1948(13) & 1.1994(35) & 1.2007(45) & 1.4880(25) & 1.4954(71) & 1.4976(93) & 0.62435(38) & 0.6253(12) & 0.6255(14)\\
0.28 & 0.674 & 1.45560(57) & 1.4559(19) & 1.4553(23) & 2.1483(15) & 2.1488(46) & 2.1463(55) & 0.649771(88) & 0.64982(34) & 0.65018(63)\\
0.285 & 0.680 & 1.64960(36) & 1.6528(18) & 1.6542(20) & 2.7412(11) & 2.7512(56) & 2.7556(66) & 0.657285(29) & 0.65753(20) & 0.65752(23)\\
0.29 & 0.744 & 1.81275(27) & 1.8146(13) & 1.8153(17) & 3.30150(96) & 3.3079(47) & 3.3107(59) & 0.660557(15) & 0.66062(11) & 0.66052(14)\\
0.295 & 0.652 & 1.95697(22) & 1.9584(14) & 1.9581(16) & 3.84240(84) & 3.8482(55) & 3.8470(61) & 0.6623272(94) & 0.662309(90) & 0.662315(99)\\
0.30 & 0.739 & 2.08782(18) & 2.08790(68) & 2.08748(72) & 4.36980(76) & 4.3701(28) & 4.3684(30) & 0.6634049(63) & 0.663396(29) & 0.663379(35)\\
\end{tabular}
\end{ruledtabular}
\end{table}

\begin{table}[H]
\caption{Observables at $L=16$.}
\label{tab:L16}
\footnotesize
\begin{ruledtabular}
\begin{tabular}{lrrrrrrrrrr}
& & \multicolumn{3}{c}{$\langle|M|\rangle$} & \multicolumn{3}{c}{$\chi/V$} & \multicolumn{3}{c}{$U_4$}\\
$\kappa$ & ESS & HMC & IS & IMH & HMC & IS & IMH & HMC & IS & IMH\\
\hline
0.25 & 0.860 & 0.15857(36) & 0.15871(95) & 0.1598(12) & 0.03923(17) & 0.03927(42) & 0.03968(54) & 0.0289(30) & 0.036(11) & 0.037(13)\\
0.255 & 0.804 & 0.19209(52) & 0.1924(12) & 0.1924(15) & 0.05717(29) & 0.05737(64) & 0.05721(82) & 0.0541(32) & 0.057(10) & 0.053(13)\\
0.26 & 0.724 & 0.25343(85) & 0.2541(10) & 0.2523(13) & 0.09753(60) & 0.09784(76) & 0.09617(91) & 0.1160(33) & 0.1255(63) & 0.1322(72)\\
0.265 & 0.556 & 0.3948(16) & 0.3968(30) & 0.4023(44) & 0.2208(15) & 0.2230(32) & 0.2283(46) & 0.2739(30) & 0.2761(78) & 0.2842(93)\\
0.2675 & 0.501 & 0.5431(22) & 0.5442(36) & 0.5458(46) & 0.3832(24) & 0.3852(42) & 0.3865(54) & 0.4067(24) & 0.4084(40) & 0.4094(53)\\
0.27 & 0.415 & 0.7611(25) & 0.7671(42) & 0.7654(51) & 0.6738(33) & 0.6842(58) & 0.6797(71) & 0.5306(15) & 0.5329(27) & 0.5340(33)\\
0.2705 & 0.387 & 0.8096(24) & 0.8182(25) & 0.8192(32) & 0.7462(33) & 0.7603(35) & 0.7619(46) & 0.5502(14) & 0.5536(14) & 0.5546(18)\\
0.2725 & 0.384 & 1.0034(19) & 1.0090(40) & 1.0103(58) & 1.0722(30) & 1.0836(69) & 1.0892(95) & 0.60801(74) & 0.6085(15) & 0.6071(22)\\
0.275 & 0.320 & 1.1963(11) & 1.1999(31) & 1.1998(34) & 1.4661(21) & 1.4746(71) & 1.4743(75) & 0.64005(28) & 0.64041(63) & 0.64030(81)\\
0.28 & 0.256 & 1.45585(41) & 1.4553(21) & 1.4525(27) & 2.1359(12) & 2.1342(55) & 2.1266(69) & 0.656864(45) & 0.65686(35) & 0.65653(54)\\
0.285 & 0.244 & 1.64930(27) & 1.6470(36) & 1.6394(95) & 2.73145(88) & 2.725(11) & 2.701(27) & 0.661280(16) & 0.66098(36) & 0.6603(11)\\
0.29 & 0.441 & 1.81248(20) & 1.8125(12) & 1.8116(15) & 3.29377(74) & 3.2938(43) & 3.2905(54) & 0.6631851(84) & 0.663184(70) & 0.663238(81)\\
0.295 & 0.486 & 1.95673(16) & 1.9567(11) & 1.9558(13) & 3.83592(64) & 3.8357(43) & 3.8323(49) & 0.6642032(51) & 0.664195(45) & 0.664202(58)\\
0.30 & 0.453 & 2.08761(14) & 2.08726(64) & 2.08723(80) & 4.36422(58) & 4.3627(26) & 4.3626(33) & 0.6648198(35) & 0.664825(22) & 0.664825(29)\\
\end{tabular}
\end{ruledtabular}
\end{table}

\begin{table}[H]
\caption{Observables at $L=20$.}
\label{tab:L20}
\footnotesize
\begin{ruledtabular}
\begin{tabular}{lrrrrrrrrrr}
& & \multicolumn{3}{c}{$\langle|M|\rangle$} & \multicolumn{3}{c}{$\chi/V$} & \multicolumn{3}{c}{$U_4$}\\
$\kappa$ & ESS & HMC & IS & IMH & HMC & IS & IMH & HMC & IS & IMH\\
\hline
0.25 & 0.799 & 0.12630(29) & 0.12716(81) & 0.1274(10) & 0.02491(11) & 0.02518(28) & 0.02519(38) & 0.0190(33) & 0.021(11) & 0.023(15)\\
0.255 & 0.765 & 0.15269(42) & 0.1538(10) & 0.1532(14) & 0.03626(19) & 0.03666(47) & 0.03618(63) & 0.0356(36) & 0.041(13) & 0.047(15)\\
0.26 & 0.629 & 0.20128(70) & 0.20137(89) & 0.2012(12) & 0.06224(40) & 0.06237(53) & 0.06217(71) & 0.0788(38) & 0.0718(75) & 0.0721(96)\\
0.265 & 0.448 & 0.3209(15) & 0.3224(27) & 0.3254(43) & 0.1501(12) & 0.1504(25) & 0.1533(42) & 0.2180(37) & 0.2176(87) & 0.213(14)\\
0.2675 & 0.316 & 0.4683(23) & 0.4681(41) & 0.4791(73) & 0.2920(23) & 0.2897(46) & 0.3012(89) & 0.3753(31) & 0.3752(60) & 0.3870(89)\\
0.27 & 0.227 & 0.7216(29) & 0.7305(49) & 0.7237(78) & 0.6039(35) & 0.6105(63) & 0.603(10) & 0.5351(19) & 0.5449(34) & 0.5391(58)\\
0.2705 & 0.192 & 0.7857(26) & 0.7831(45) & 0.7843(87) & 0.6942(34) & 0.6888(50) & 0.6896(90) & 0.5623(15) & 0.5632(36) & 0.5646(68)\\
0.2725 & 0.173 & 1.0056(18) & 1.0085(59) & 1.0062(78) & 1.0562(28) & 1.0606(92) & 1.058(11) & 0.62406(64) & 0.6261(23) & 0.6250(31)\\
0.275 & 0.200 & 1.19885(83) & 1.2039(30) & 1.2005(38) & 1.4591(17) & 1.4681(69) & 1.4600(85) & 0.64905(19) & 0.65092(55) & 0.65051(65)\\
0.28 & 0.347 & 1.45633(32) & 1.45810(93) & 1.4593(11) & 2.13123(91) & 2.1357(26) & 2.1390(30) & 0.660360(25) & 0.660745(90) & 0.66079(10)\\
0.285 & 0.453 & 1.64949(21) & 1.6518(12) & 1.6507(14) & 2.72799(69) & 2.7353(38) & 2.7319(47) & 0.6632100(97) & 0.663372(71) & 0.663258(89)\\
0.29 & 0.488 & 1.81256(16) & 1.8140(11) & 1.8138(11) & 3.29093(58) & 3.2958(39) & 3.2954(41) & 0.6644344(52) & 0.664486(69) & 0.664498(50)\\
0.295 & 0.478 & 1.95678(13) & 1.9574(13) & 1.9582(12) & 3.83354(51) & 3.8360(49) & 3.8392(46) & 0.6650878(32) & 0.665072(73) & 0.665064(75)\\
0.30 & 0.571 & 2.08767(11) & 2.08804(39) & 2.08847(50) & 4.36227(46) & 4.3637(16) & 4.3655(21) & 0.6654830(22) & 0.665515(16) & 0.665514(17)\\
\end{tabular}
\end{ruledtabular}
\end{table}

\begin{table}[H]
\caption{Observables at $L=24$.}
\label{tab:L24}
\footnotesize
\begin{ruledtabular}
\begin{tabular}{lrrrrrrrrrr}
& & \multicolumn{3}{c}{$\langle|M|\rangle$} & \multicolumn{3}{c}{$\chi/V$} & \multicolumn{3}{c}{$U_4$}\\
$\kappa$ & ESS & HMC & IS & IMH & HMC & IS & IMH & HMC & IS & IMH\\
\hline
0.25 & 0.752 & 0.10537(25) & 0.10479(77) & 0.10481(99) & 0.017387(79) & 0.01736(26) & 0.01737(32) & 0.0124(35) & -0.011(18) & -0.005(19)\\
0.255 & 0.658 & 0.12744(36) & 0.12682(98) & 0.1269(13) & 0.02538(14) & 0.02534(40) & 0.02533(51) & 0.0208(40) & 0.003(18) & 0.009(22)\\
0.26 & 0.522 & 0.16815(60) & 0.16746(83) & 0.1674(18) & 0.04378(30) & 0.04365(42) & 0.0434(10) & 0.0523(44) & 0.0467(95) & 0.051(12)\\
0.265 & 0.302 & 0.2717(14) & 0.2675(27) & 0.2660(39) & 0.10945(97) & 0.1068(19) & 0.1069(27) & 0.1774(43) & 0.173(13) & 0.158(18)\\
0.2675 & 0.166 & 0.4052(25) & 0.3976(47) & 0.3962(69) & 0.2255(22) & 0.2182(43) & 0.2180(66) & 0.3311(40) & 0.328(11) & 0.321(14)\\
0.27 & 0.041 & 0.6820(33) & 0.656(15) & 0.665(16) & 0.5421(38) & 0.512(15) & 0.521(18) & 0.5317(23) & 0.516(12) & 0.522(13)\\
0.2705 & 0.056 & 0.7510(31) & 0.7518(81) & 0.7516(86) & 0.6362(37) & 0.636(12) & 0.636(12) & 0.5619(19) & 0.5627(36) & 0.5618(41)\\
0.2725 & 0.063 & 0.9988(18) & 0.9946(98) & 0.989(16) & 1.0340(27) & 1.024(14) & 1.019(22) & 0.63129(64) & 0.6313(44) & 0.6273(75)\\
0.275 & 0.080 & 1.19878(69) & 1.2040(29) & 1.2082(32) & 1.4523(15) & 1.4618(67) & 1.4715(75) & 0.65392(15) & 0.65592(46) & 0.65625(53)\\
0.28 & 0.119 & 1.45561(28) & 1.4564(14) & 1.4554(14) & 2.12604(79) & 2.1278(39) & 2.1251(40) & 0.662207(18) & 0.66247(11) & 0.66252(11)\\
0.285 & 0.316 & 1.64910(18) & 1.6479(11) & 1.6476(14) & 2.72451(60) & 2.7206(34) & 2.7195(45) & 0.6642475(67) & 0.664269(49) & 0.664242(68)\\
0.29 & 0.352 & 1.81228(14) & 1.81218(82) & 1.8121(11) & 3.28823(50) & 3.2877(30) & 3.2872(41) & 0.6651096(36) & 0.665166(28) & 0.665193(35)\\
0.295 & 0.399 & 1.95656(11) & 1.95643(75) & 1.95605(98) & 3.83130(43) & 3.8307(29) & 3.8292(38) & 0.6655675(22) & 0.665604(22) & 0.665593(30)\\
0.30 & 0.333 & 2.087490(94) & 2.08718(46) & 2.08728(56) & 4.36032(39) & 4.3590(19) & 4.3594(23) & 0.6658434(15) & 0.665858(12) & 0.665857(13)\\
\end{tabular}
\end{ruledtabular}
\end{table}

\begin{table}[H]
\caption{Observables at $L=32$.}
\label{tab:L32}
\footnotesize
\begin{ruledtabular}
\begin{tabular}{lrrrrrrrrrr}
& & \multicolumn{3}{c}{$\langle|M|\rangle$} & \multicolumn{3}{c}{$\chi/V$} & \multicolumn{3}{c}{$U_4$}\\
$\kappa$ & ESS & HMC & IS & IMH & HMC & IS & IMH & HMC & IS & IMH\\
\hline
0.25 & 0.612 & 0.07859(19) & 0.07917(56) & 0.07937(79) & 0.009685(45) & 0.00989(13) & 0.00990(19) & 0.0036(37) & -0.012(14) & -0.003(18)\\
0.255 & 0.475 & 0.09495(28) & 0.09555(79) & 0.0960(11) & 0.014121(80) & 0.01439(23) & 0.01446(32) & 0.0083(43) & -0.008(15) & 0.005(21)\\
0.26 & 0.313 & 0.12509(48) & 0.12510(78) & 0.1257(11) & 0.02442(18) & 0.02451(31) & 0.02473(39) & 0.0267(50) & 0.026(12) & 0.017(15)\\
0.265 & 0.104 & 0.2035(12) & 0.2004(39) & 0.2035(56) & 0.06299(66) & 0.0618(24) & 0.0638(34) & 0.1114(55) & 0.071(27) & 0.069(32)\\
0.2675 & 0.051 & 0.3168(24) & 0.3044(69) & 0.310(10) & 0.1435(18) & 0.1332(55) & 0.1345(78) & 0.2588(54) & 0.229(22) & 0.270(34)\\
0.27 & 0.021 & 0.6267(40) & 0.625(13) & 0.613(15) & 0.4604(44) & 0.453(13) & 0.437(16) & 0.5275(31) & 0.536(10) & 0.533(12)\\
0.2705 & 0.013 & 0.7155(38) & 0.708(12) & 0.705(16) & 0.5739(45) & 0.561(17) & 0.560(17) & 0.5684(23) & 0.5683(73) & 0.567(10)\\
0.2725 & 0.0069 & 0.9992(21) & 0.995(11) & 1.0036(91) & 1.0214(29) & 1.006(22) & 1.021(19) & 0.64345(82) & 0.6476(19) & 0.6501(14)\\
0.275 & 0.033 & 1.19958(48) & 1.2031(37) & 1.2065(30) & 1.4473(11) & 1.4543(88) & 1.4625(70) & 0.659393(63) & 0.66042(33) & 0.66060(32)\\
0.28 & 0.083 & 1.45572(21) & 1.4579(11) & 1.4585(13) & 2.12318(60) & 2.1292(31) & 2.1308(37) & 0.6641477(94) & 0.664370(48) & 0.664428(78)\\
0.285 & 0.162 & 1.64911(14) & 1.6514(10) & 1.6514(14) & 2.72235(45) & 2.7299(33) & 2.7301(45) & 0.6653037(37) & 0.665349(31) & 0.665311(54)\\
0.29 & 0.253 & 1.81231(10) & 1.81324(69) & 1.81500(89) & 3.28664(38) & 3.2900(25) & 3.2963(32) & 0.6657907(20) & 0.665796(19) & 0.665819(24)\\
0.295 & 0.266 & 1.956606(84) & 1.95728(61) & 1.95722(88) & 3.83009(33) & 3.8327(24) & 3.8326(34) & 0.6660485(13) & 0.666047(13) & 0.666020(20)\\
0.30 & 0.241 & 2.087515(71) & 2.08769(35) & 2.08781(44) & 4.35923(30) & 4.3599(14) & 4.3604(19) & 0.6662038(9) & 0.6662140(89) & 0.666215(14)\\
\end{tabular}
\end{ruledtabular}
\end{table}

\end{document}